\documentclass[3p,times,procedia]{elsarticle}
\usepackage{nupha_ecrc}

\volume{00}

\firstpage{1}

\journalname{Nuclear Physics A}

\runauth{Leticia Cunqueiro}

\jid{nupha}

\jnltitlelogo{Nuclear Physics A}

\usepackage{amssymb}

\usepackage[figuresright]{rotating}

\begin{document}

\begin{frontmatter}




\title{Jet Production at RHIC and LHC}


\author{Leticia Cunqueiro}

\address{University of M\"unster and CERN}

\begin{abstract}
Recent results on jet production in heavy ion collisions at RHIC and
the LHC are discussed, with emphasis on inclusive jet yields and
semi-inclusive hadron-triggered and vector boson-triggered recoil jet
yields as well as their azimuthal angular correlations. I will also
discuss the constraints that these observables impose on the opacity of the medium, the flavour dependence of
energy loss, the interplay of perturbative and non perturbative
effects and the change of the degrees of freedom of the medium with
the resolution of the probe.
\end{abstract}




\end{frontmatter}





\section{Introduction}
\label{intro}

 The scope of the heavy ion jet physics program at RHIC and LHC  is to understand the
behaviour of QCD matter at the limit of high density and temperature via the
study of the dynamics of the jet-medium interactions.

 Jet physics in heavy ion collisions is a multiscale problem. Hard
scales govern the perturbative production of the elementary scattering
and subsequent branching in vacuum and in medium. While jet
constituents may interact strongly with the medium at scales corresponding to the 
temperature of the plasma. 

 The characterization of medium modifications of jet distributions
benefits from observables that are well-defined, that preserve the infrared
and collinear safety of the measurement and thus allow for
 a direct connection to the theory. It also requires the control of
 the large combinatorial background present in heavy ion collisions.

The first generation of jet measurements are the jet production cross
sections and their suppression relative to the vacuum proton-proton
reference. These ``disappeareance'' measurements indicate that a
significant amount of energy is radiated out of the jet area but do
not impose severe constrains on the dynamics of jet-medium
interactions. 
 
The second generation of observables are semi-inclusive jet rates, using
hadrons, jets or vector-bosons as triggers.  Coincidence measurements
allow exploration of interjet broadening and inspection of low jet
momenta and high resolution $R$. Vector-boson
triggers allow for a quantification of the energy lost by the
recoiling jet. 

The third generation of observables are jet shapes. Possible
modifications of the intrajet distributions are currently explored via the jet
mass \cite{Acharya:2017goa}, dispersion $p_{T}D$, angularities \cite{Cunqueiro:2015dmx}, fragmentation functions
  \cite{ATLAS:2017iya} and jet-track correlations
  \cite{CMS:2015tla}. The role of color coherence in medium \cite{CasalderreySolana:2012ef} is
  explored by new observables like 2-subjetiness
  \cite{Zardoshti:2017yiy}, and  soft drop subjet momentum
  balance \cite{CMS:2016jys}.  Those two measurements were designed to
  understand whether the subjet structure is resolved 
  by the medium, depending on the angular scale, and consequently,
  determine whether subjets interact in the medium
  independently or coherently. The second and
  third generation observables can be combined in the measurement of
  substructure of recoiling jets, as it is done in
  \cite{Zardoshti:2017yiy}.  Note as well that this third generation
  of jet observables won't be discussed here since they were the subject
  of another talk \cite{Marta}.

Open questions include the flavour dependence of
energy loss, the dependence of energy loss on jet substructure, the
role of color-coherence effects, the interplay between weak
and strong coupling effects, the role of the medium response or
correlated background and the change with the probe resolution scale of the
medium degrees of freedom. 

\section{Inclusive jet suppression}

 Inclusive jet yields have been
measured in Pb-Pb collisions at the LHC over a wide kinematic range
from a few tens of GeV to the TeV scale.  The suppression of such yields relative to the  binary
collision-scaled proton-proton reference, is flat up to jet transverse momentum $p_{T}$ of 1 TeV
for jet resolution $R=0.4$.  The strong suppression of $1$ TeV jets is
a striking observation, given the short size of the medium compared to
the hadronization length of the shower of such energetic probes. 

\begin{figure}[h]
\centering
\includegraphics[width=0.45\textwidth]{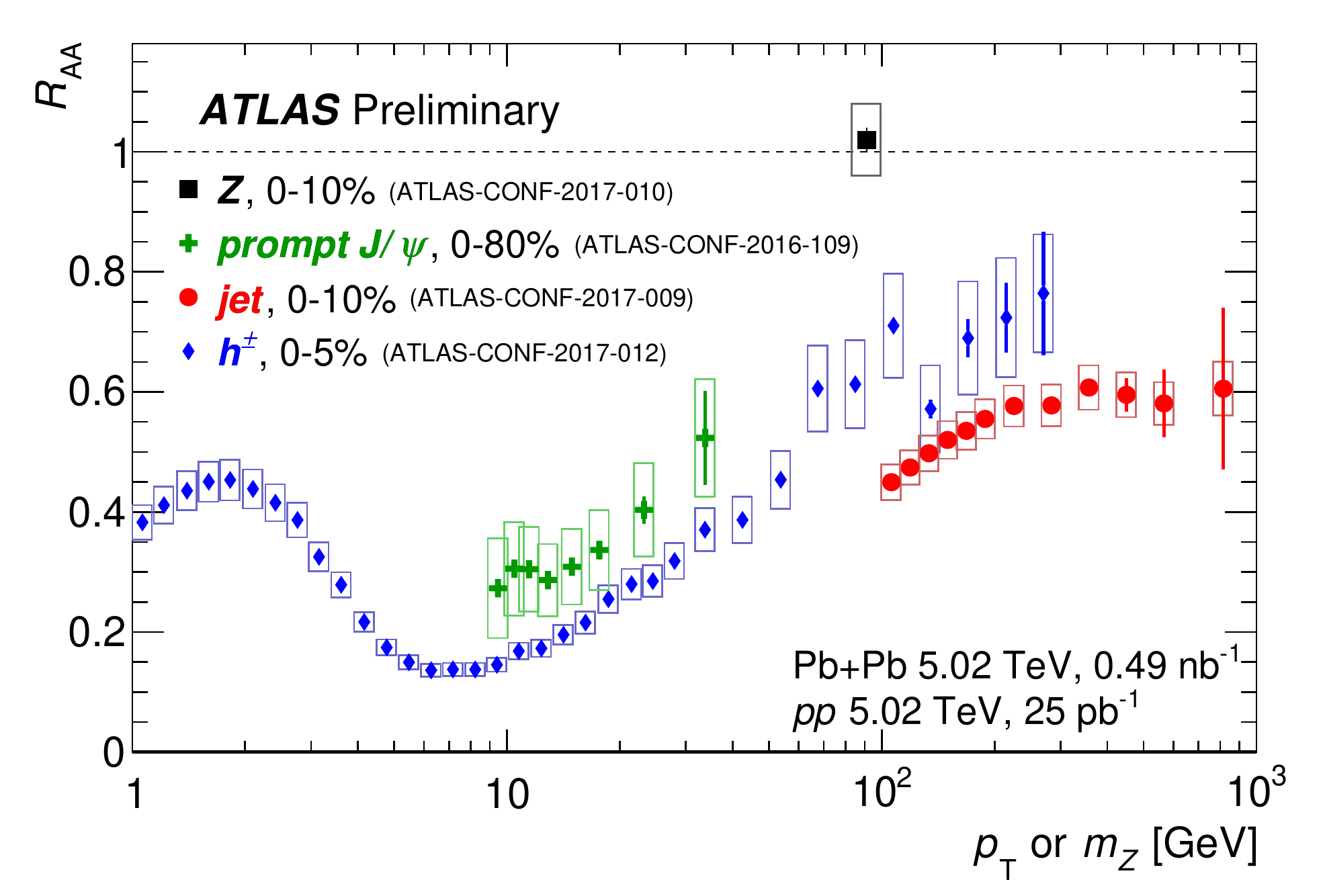}
\includegraphics[width=0.45\textwidth]{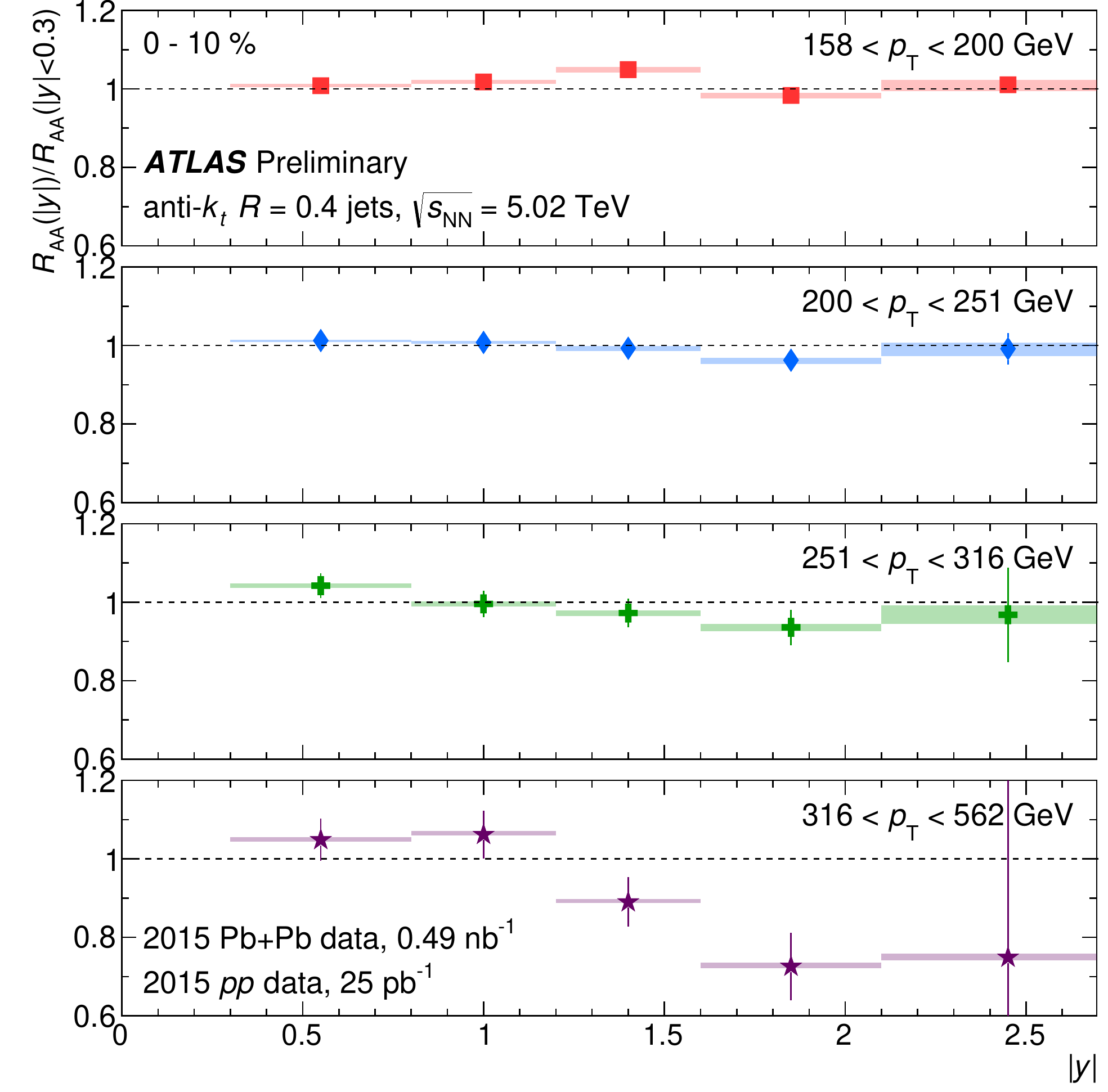}
\includegraphics[width=0.55\textwidth]{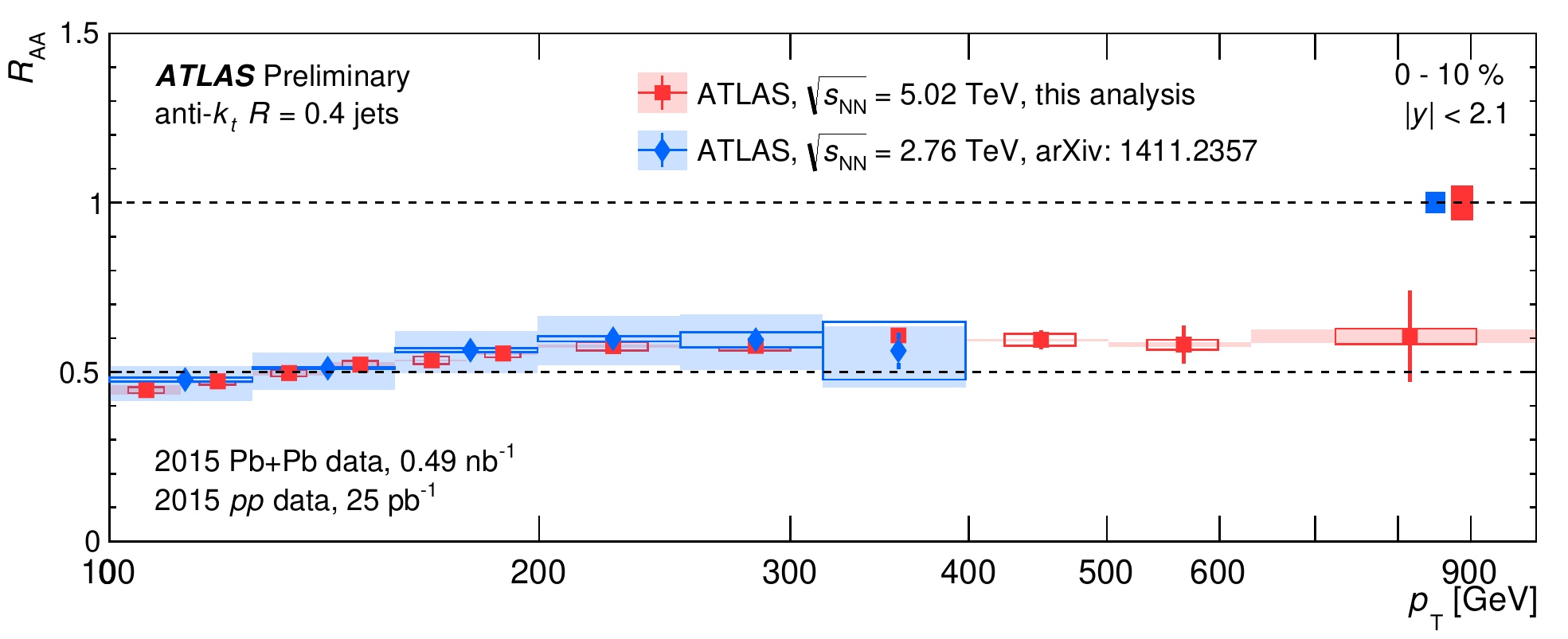}
\caption{ATLAS results for hadron and jet nuclear modification factor
  in central Pb-Pb collisions at 5.02 TeV, along with rapidity dependence}
\label{fig:RAAAtlas}
\end{figure}

Fig. \ref{fig:RAAAtlas} (top left), shows ATLAS results for inclusive jet and hadron nuclear
modification factor in Pb-Pb collisions at 5.02 TeV \cite{ATLAS:2017wvp}. Data for
prompt $J/\psi$ and $Z$ bosons are also shown. In
Fig. \ref{fig:RAAAtlas} (bottom), the jet
suppression is magnified and compared to 2.76 TeV results. There is no
evidence of collision energy dependence. It should be noted that the magnitude of the
suppression $R_{AA}$ is not a direct measure of the opacity of the
medium to high $p_{T}$ probes, but rather depends on other elements such
the the slope of the underlying hard parton spectrum and the
quark/gluon fractions. A different medium opacity at the two
colliding energies could manifest itself in a comparable $R_{AA}$. 

The measurement of jet yields at forward rapidities is complementary
to studying the $\sqrt{s}$ dependence, since it also offers the possibility to vary the
medium density, the quark/gluon content and the spectral slope. It was
predicted \cite{Renk:2014gwa}  that at forward rapidities and close to the
kinematic limit where the spectrum becomes steep, the suppression of
forward jets will increase relative to midrapidity. The ATLAS
measurement, see Fig. \ref{fig:RAAAtlas} (top right), confirms this prediction qualitatively. At forward rapidity and high jet energy, the jet yields are suppressed relative
to midrapidity by 30$\%$. 

Another interesting question that rises when
exploring jet production at the TeV scale is the role of nuclear
effects. At midrapidity, for $E=1$ TeV and $\sqrt{s_{NN}}= 5$ TeV,
the parton momentum fraction is $x_{t}=2E/\sqrt{s_{NN}} =0.4$ and EMC effects may emerge.

\section{Semi-inclusive recoil jet yields and momentum balance }
\label{pp}
 Semi-inclusive trigger-jet coincidence measurements provide
several advantages:
\begin{itemize}
 \item Analysis based on semi-inclusive coincidence measurements
 allow for a precise subtraction of the background uncorrelated to the
 hard scattering, without imposing fragmentation bias on the jet
 population. When such background is removed, full correction of the yields can
  be achieved down to low jet $p_{T}$ and large $R$. 

\item When the trigger is a vector boson that does not interact
  strongly with the medium, the momentum imbalance between the trigger
  and the recoil jet provides a direct measurement of the energy lost out of the jet
  area.

\item Different choices of trigger bias towards different recoil jet
  flavour, allowing for the exploration of the flavour dependence of
  energy loss. The different trigger choices also vary the geometric bias.

\item Semi-inclusive measurements are self-normalized and as a
  consequence they don't need an interpretation of event activity  in
  terms of geometry. This is an advantage when studying energy loss in
  small systems as was shown \cite{Filip}. 
\end{itemize}

\begin{figure}[h]
\centering
\includegraphics[width=0.37\textwidth]{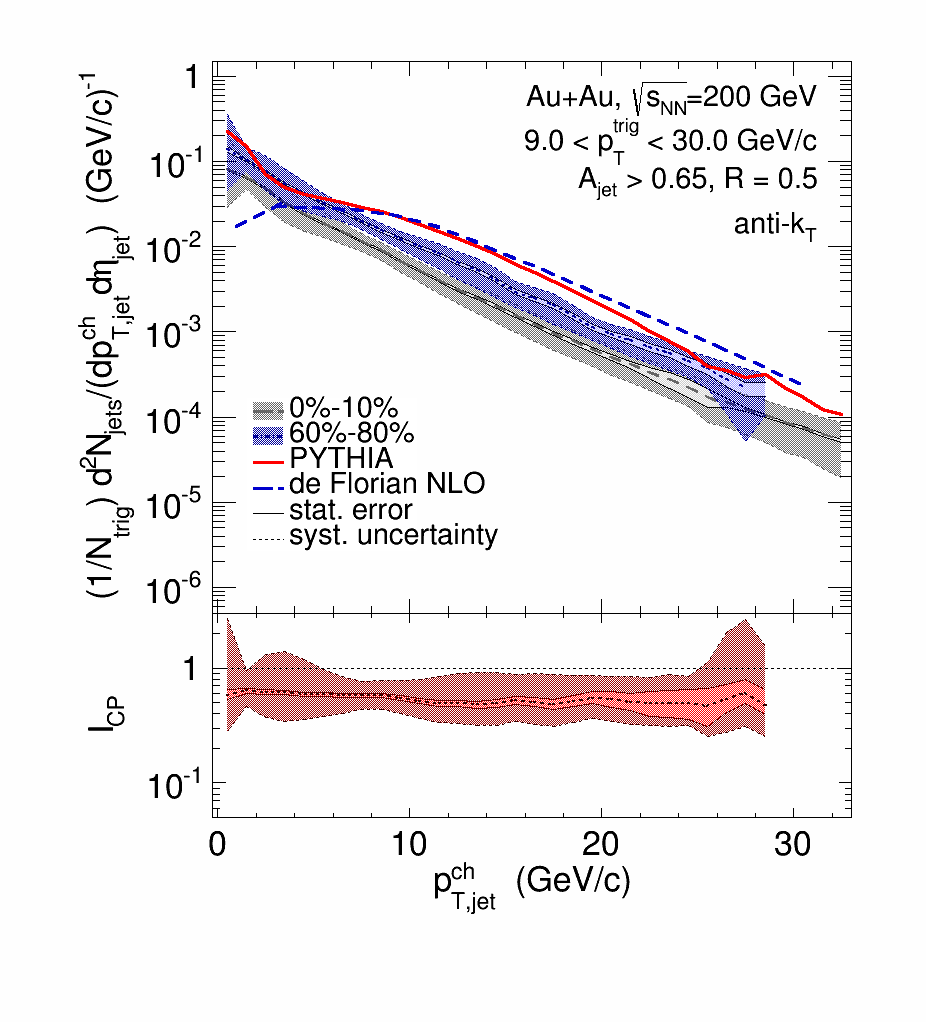}
\includegraphics[width=0.37\textwidth]{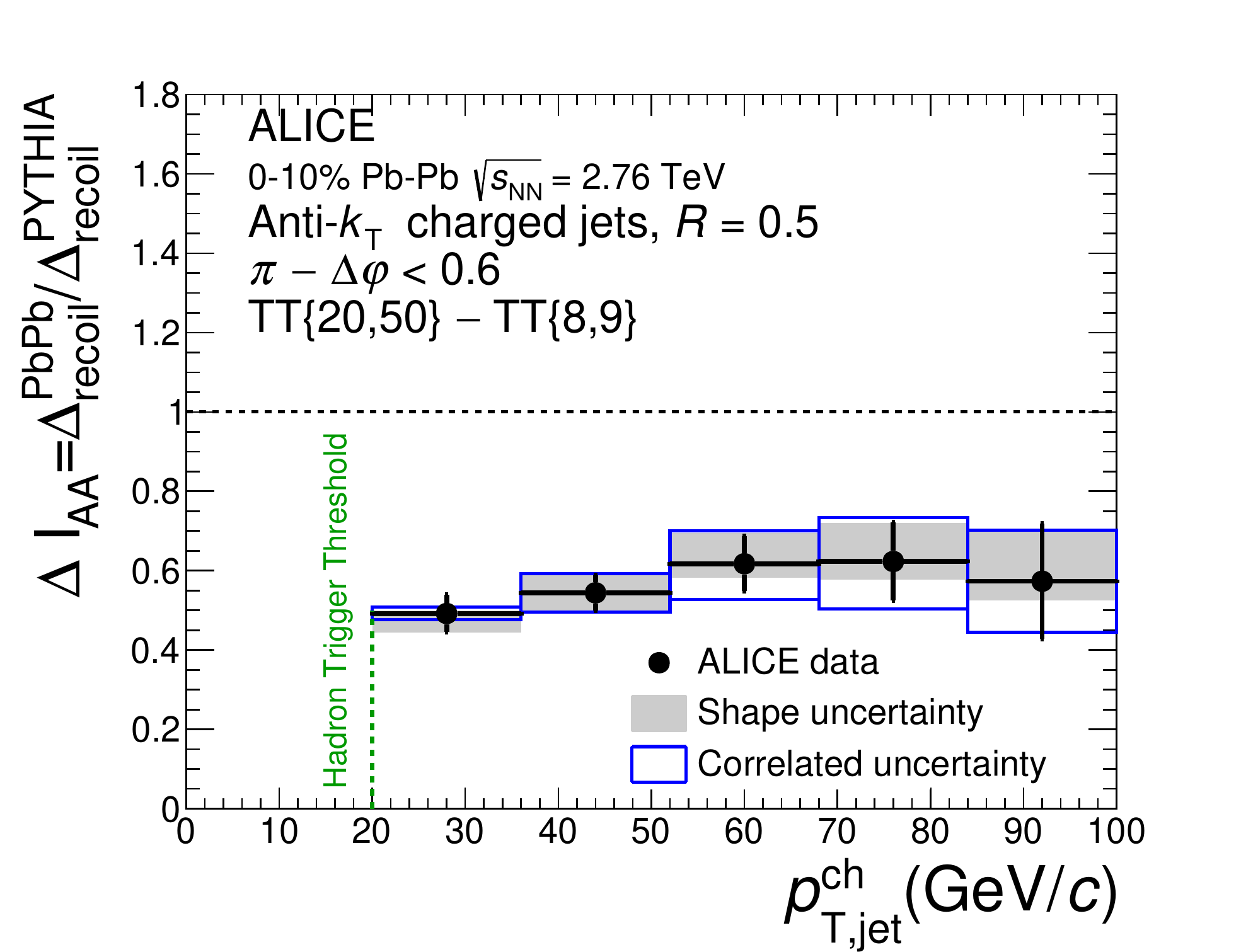}
\includegraphics[width=0.37\textwidth]{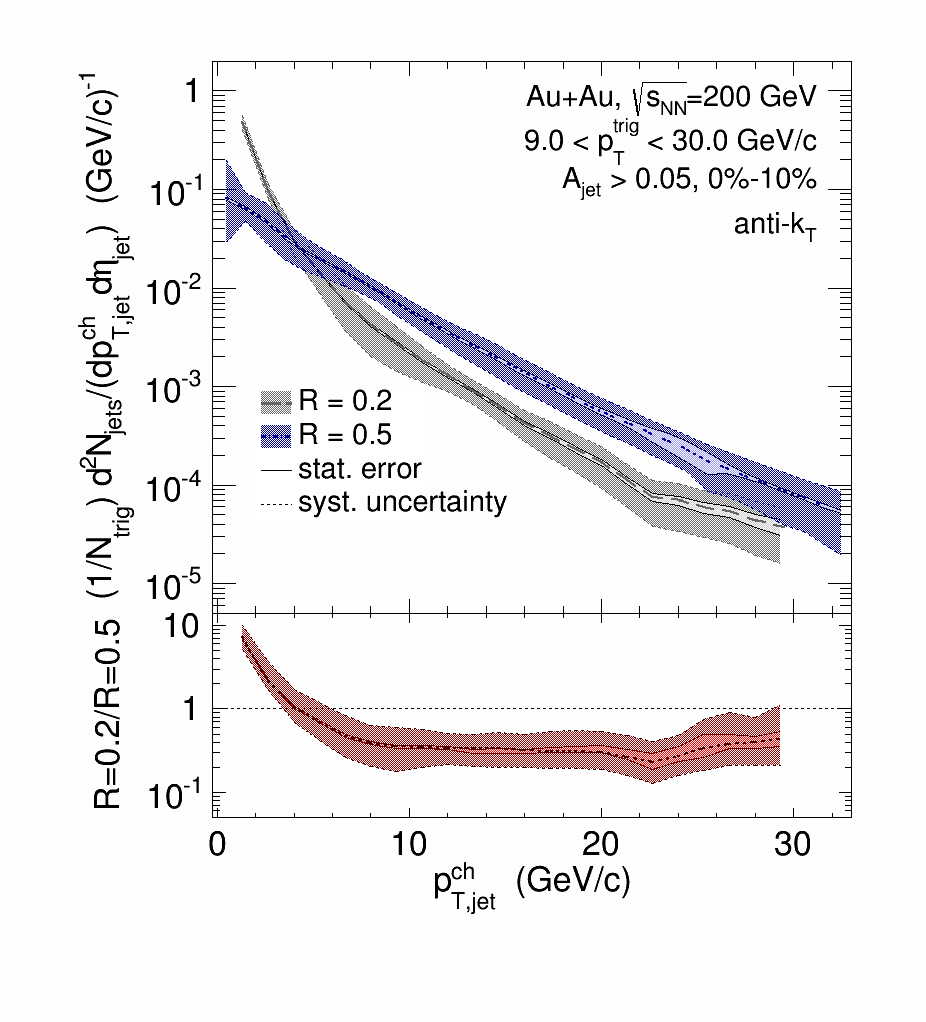}
\includegraphics[width=0.37\textwidth]{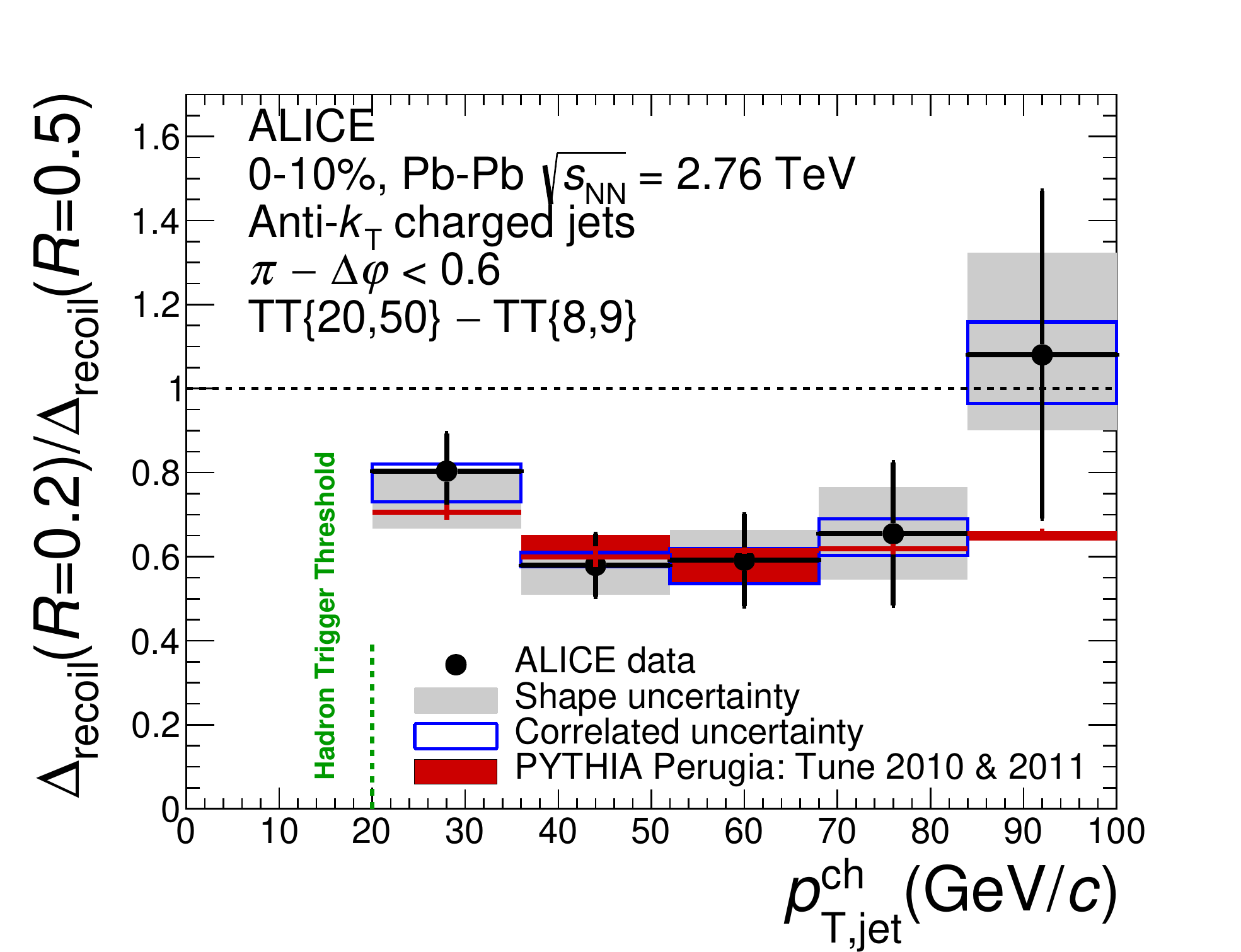}
\caption{Top: STAR semi-inclusive charged jet yields recoiling from a high
  $p_{T}$ track in central and peripheral collisions at $\sqrt(s)=$200
  GeV and their ratio $I_{CP}$ (left). Ratio of ALICE $\Delta_{Recoil}$
  observable in central relative to PYTHIA at $\sqrt(s)=$2.76 TeV
  (right). Bottom: Ratio of recoil jet distributions measured with $R=0.2$
  relative to those measured with $R=0.5$. STAR (left) and ALICE
  (right) results are shown}
\label{fig:recoilICP}
\end{figure}

The STAR collaboration has reported the semi-inclusive yields of track-based jets
recoiling from high $p_{T}$ hadrons ($9<p_{T}^{trig}<30$ GeV) \cite{Adamczyk:2017yhe}.
A key aspect of the analysis is that uncorrelated background is removed
via event mixing and the fully corrected jet yields are measured down to nearly zero $p_{T}$
for $R$ up to $0.5$. The central and peripheral corrected yields are shown in
 Fig. \ref{fig:recoilICP} (top left) together with a PYTHIA and
a NLO calculation, and their ratio, $I_{CP}$ is shown in the lower panel of the
same figure. 

ALICE has also measured the distribution of track-based jets recoiling from a high
$p_{T}$ trigger track \cite{Adam:2015doa}. To remove combinatorial background, the difference of
two exclusive trigger track classes is taken and a new observable is
defined $\Delta_{Recoil}$. The trigger track class subtraction is
proven to be effectively equivalent to the event mixing except for the
treatment of multiple
parton interactions at low $p_{T}$,  MPIs, that
are not subtracted in the latter  \cite{Adamczyk:2017yhe}.   The ratio of central Pb-Pb $\Delta_{Recoil}$ to
PYTHIA $\Delta_{Recoil}$ is shown in
Fig. \ref{fig:recoilICP} (top right) down to $p_{T}=20$ GeV and $R=0.5$. 

In Fig. \ref{fig:recoilICP} (bottom), the ratio of the recoil jet
distributions measured with $R=0.2$ relative to those measured with
$R=0.5$ is shown both by STAR (left) and ALICE (right). ALICE data reveals no
in-medium redistribution of energy within $R=0.5$ compared to the
vacuum reference. 

The finite suppression of the recoil jet distributions in
Fig. \ref{fig:recoilICP} together with the low infrared cut-off of
these measurements, indicates that the medium-induced energy loss
arises predominantly from radiation at angles larger than 0.5 relative
to the jet axis. The lost energy is reflected in the magnitude of the
spectrum shift under the assumption of negligible trigger track energy loss and it is estimated to be $8 \pm
2$ GeV at ALICE, in the jet momentum range of 60 to 100 GeV while STAR
results indicate approximately 4 GeV in the jet momentum range of 10 to 20 GeV.

\begin{figure}[h]
\centering
\includegraphics[width=0.47\textwidth]{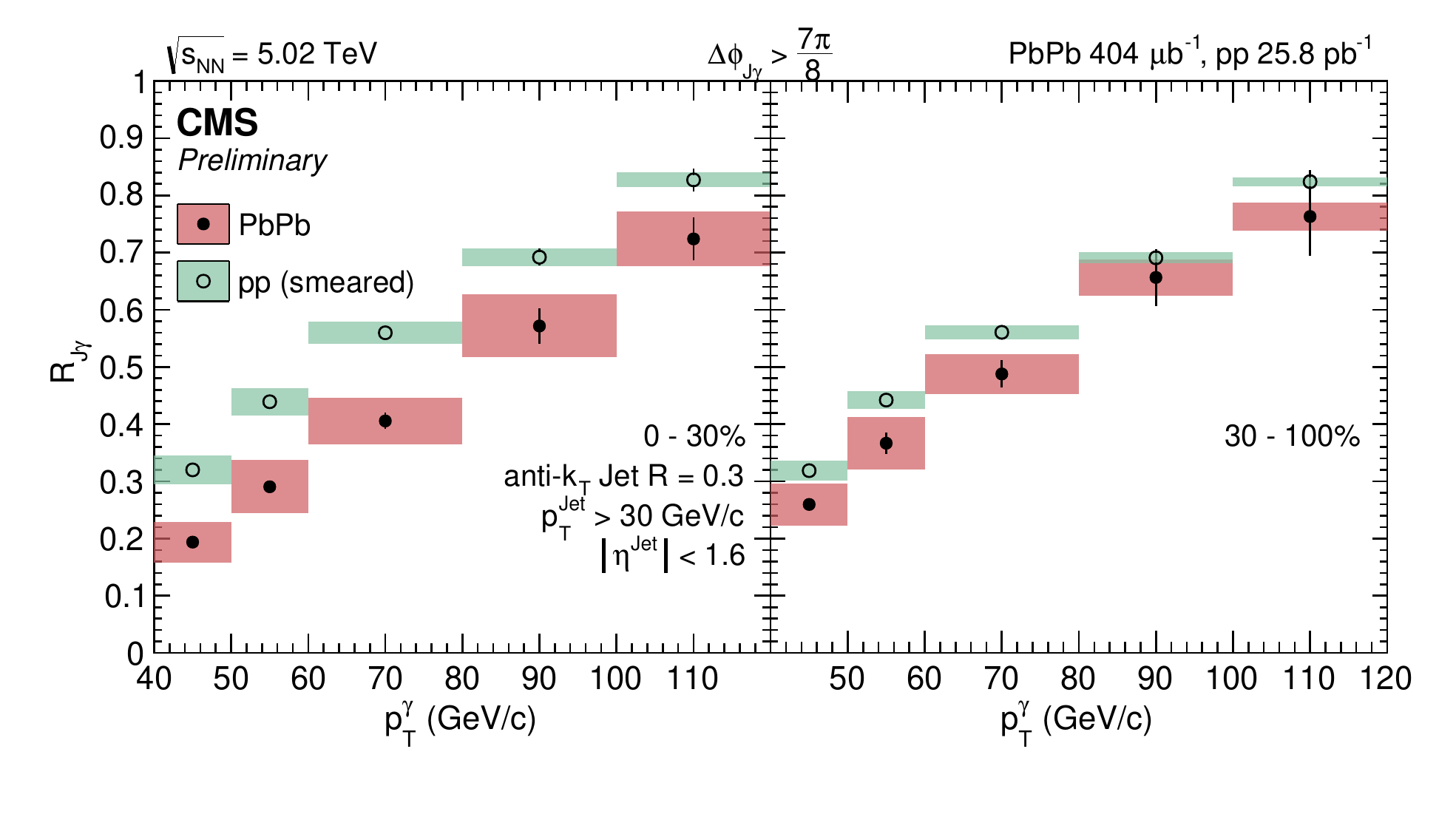}
\includegraphics[width=0.47\textwidth]{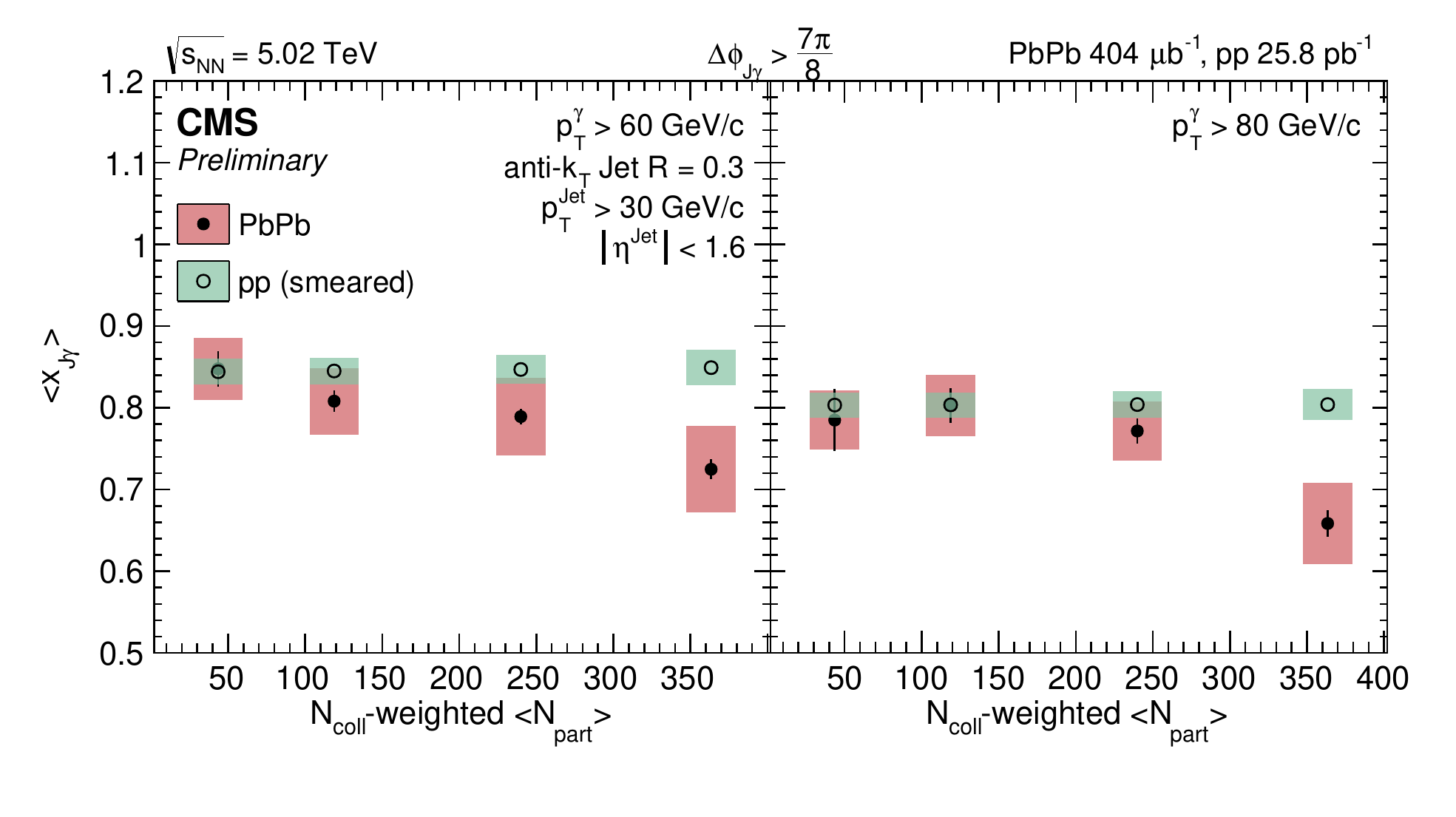}
\includegraphics[width=0.4\textwidth, height=0.35\textwidth]{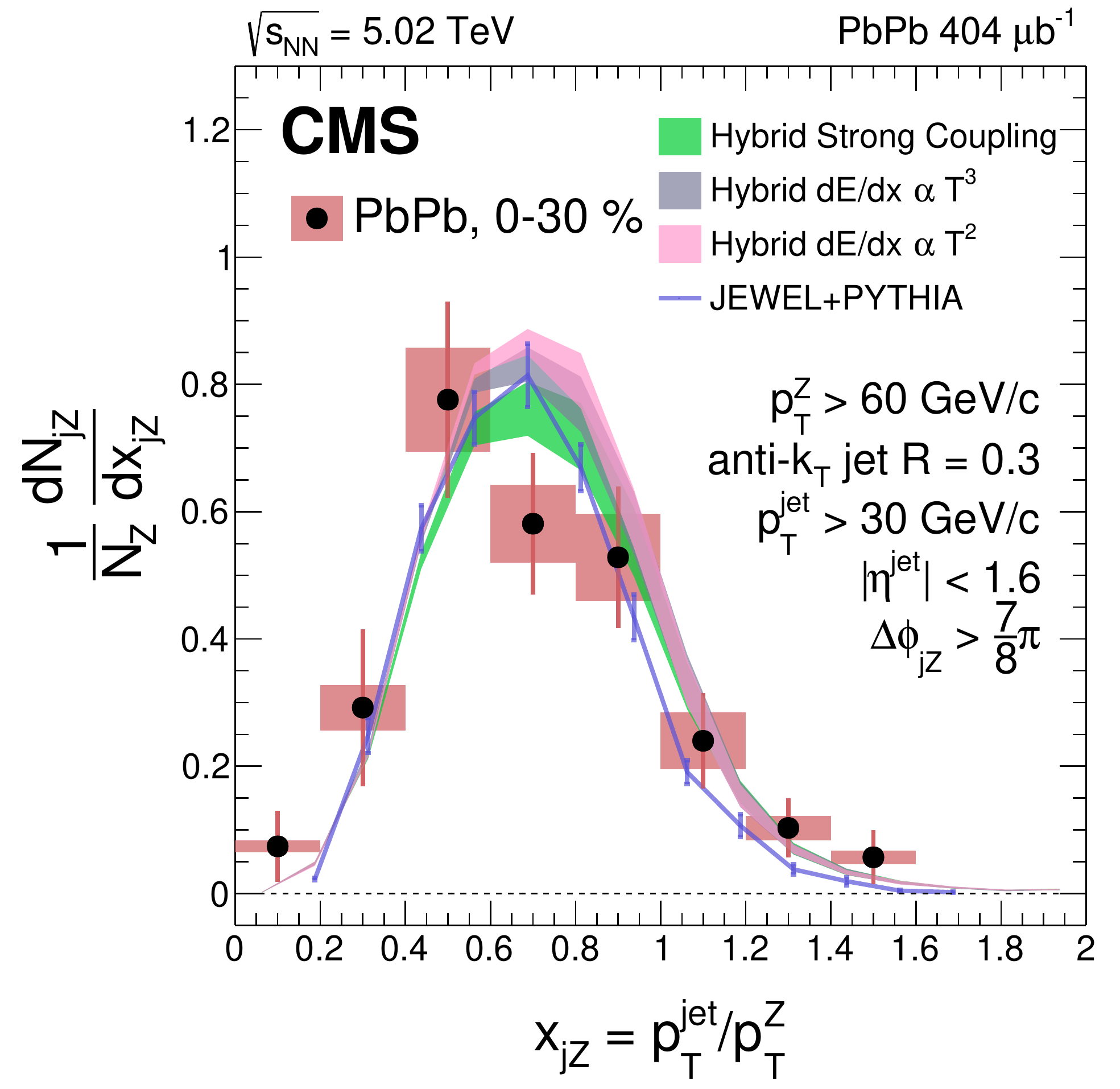}
\includegraphics[width=0.4\textwidth, height=0.35\textwidth]{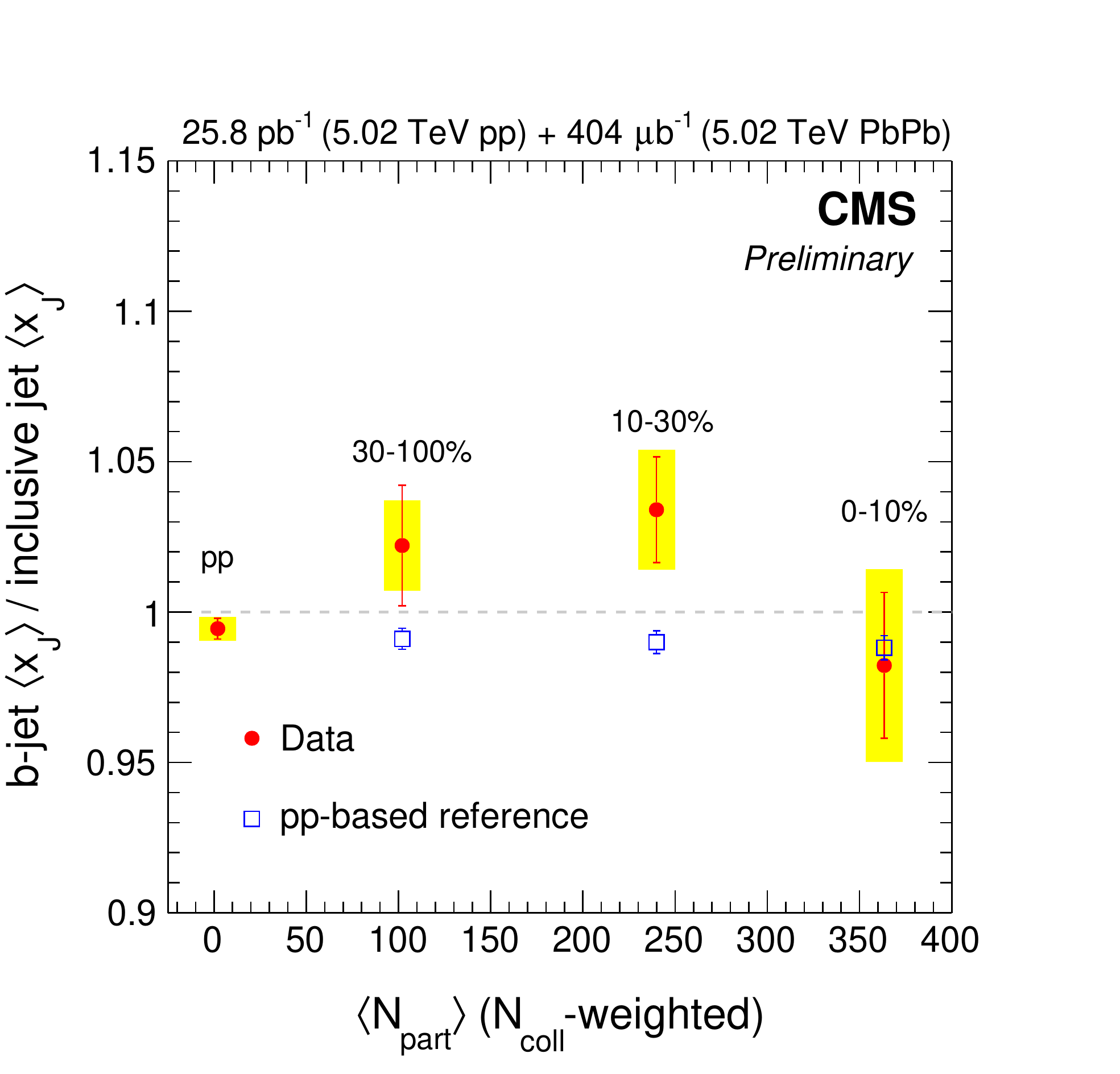}
\caption{$R_{J\gamma}$, average momentum imbalance $\angle
  x_{J\gamma} \rangle$,
  the $x_{JZ}$ distribution in central Pb-Pb events and the ratio of
  beauty dijet and dijet momentum imbalance versus centrality.  }
\label{fig:CMSrecoil}
\end{figure}

CMS and ATLAS have measured jet production in
association with isolated $\gamma$ and $Z$ bosons in proton-proton and
Pb-Pb collisions \cite{ATLAS:2016tor,ATLAS:2017zkv,CMS:2016ynj,Sirunyan:2017jic}.  The $\gamma$ and $Z$ triggers have $40<p_{T}^{\gamma}<60$ and $p_{T}^{Z}>60$ respectively, while
the recoiling jets have $p_{T}^{jet}> 30$ GeV with $R=0.3$ and $R=0.4$, respectively. 
The CMS measurement of $R_{J\gamma}$ , the number of jet coincidences per $\gamma$ trigger,
is shown in Fig. \ref{fig:CMSrecoil} (top left). The central Pb-Pb data
is below the vacuum reference (proton-proton results smeared by
background fluctuations) indicating that a
significant fraction of the recoil jets lose energy and their momentum is
shifted below the 30 GeV threshold. The hardening of the recoil jet
yield with increasing photon momentum is reflectedin the increasing
trend of  $R_{J\gamma}$.

The momentum asymmetry $x_{JB}$, the ratio of the jet and
boson momentum, would be a
$\delta$ function only at LO. Higher-order corrections broaden its
distribution. A shift in the mean of this distribution is used to
quantify the average energy lost by the recoil jets.   

Fig. \ref{fig:CMSrecoil} (top right) shows the average fraction of the photon
momentum carried by the recoiling jets as a function of centrality. One
can see that in the most central bin, the average
shift between Pb-Pb data and the smeared pp reference is compatible
with $\approx 10 $ GeV of energy radiated out of the jet area. 
In Fig.  \ref{fig:CMSrecoil} (bottom left) the asymmetry distribution (in this case for Z-triggered
recoiled jets) is shown for central Pb-Pb collisions compared to
several theory models. It should be noted that the data points are not
fully corrected (not unfolded for background fluctuations and detector
effects) so theory predictions are consistently smeared with parametrizations of the
response provided by the experiment.   
 
 In the absense of a consistent theoretical model incorporating
weakly and strongly coupled energy loss, both limits are compared
independently to data. JEWEL \cite{Zapp:2013vla} is a Monte Carlo model that considers elastic and radiative
energy loss in the medium in a weakly coupled framework. In the Hybrid
model \cite{Casalderrey-Solana:2015vaa} parton showers are 
generated by Pythia and the interactions with the medium are
implemented by changing the momenta of the partons using an
analytical calculation of the lost energy according to strong coupling expectations from string calculations in the
gauge/gravity duality. Its prediction is the green curve in plot. To explore the sensitivity of the observable to the details of
energy loss, two parametric perturbative limits for the energy loss
are also considered, one proportional to the third
power of the temperature, as expected from radiative losses, one
proportional to the squared of the temperature, as expected from
collisional losses. One can see that the differences between the
different approaches are not significant enough to discriminate between the
details of the energy loss process as implemented in the models.  Other
purely perturbative calculations\cite{Kang:2017xnc} reproduce
boson-tagged jet data farily well. 

$Z$ and $\gamma$ triggers enhance the sample of quark recoiled jets
compared to jet triggers, so these data may illuminate the
parton-flavour dependence of energy loss. 

At high jet $p_{T}$ where quark
mass effects are negligible, a b-jet is essentially a quark jet. 
Triggering on b-dijets is interesting because the requirement of a large azimuthal gap between
the dijet system suppresses the contribution to b-jet production of
gluon splitting. Such a contribution confuses the flavour and color
charge of the object propagating in the medium. The beauty dijet sample is consequently a cleaner measurement
of prompt heavy quark flavour than inclusive beauty jet $R_{AA}$.
In Fig. \ref{fig:CMSrecoil} (bottom right) one can see that the momentum imbalance
of the b dijet system is consistent with that of the inclusive dijet system
across all centrality bins for trigger and recoil jets of $p_{T} > 100$
and $p_{T} > 40$ GeV respectively.  Since the jet-triggered recoil jet
population is expected to be gluon-dominated, the previous plot points to
no significant differences in quark
and gluon energy loss in the given kinematic regime. 
\section{Medium-induced acoplanarity}
\label{PbPb}

The combined analysis of jet energy loss and momentum
broadening can constrain the underlying mechanism of energy loss.  
In the perturbative BDMPS formalism \cite{Baier:1996sk}, for instance, energy loss and
momentum broadening are linearly coupled by a single parameter, the
transport coefficient $\hat{q}$. On the contrary, in the limit of
strong coupling \cite{Casalderrey-Solana:2014wca}, energy loss and broadening are uncorrelated. 

Momentum broadening can be studied experimentally via
intrajet shapes sensitive to the redistribution of jet momentum and
constituents to wider angles \cite{Kurkela:2014tla} and it can be studied as a change in the
jet direction as a whole \cite{Chen:2016vem} Here we focus on the second approach and we
inspect momentum broadening via the azimuthal angular correlation of dijet systems.

\subsection{Broadening of the primary peak of the azimuthal angular correlation}
 In a purely perturbative framework, there are two ingredients that contribute to the azimuthal
decorrelation of the recoil jet: one is the vacuum soft and collinear
radiation and other is the medium-induced effects. The latter are
described as random kicks of momentum at each scattering of the
partonic projectile with the medium. The multiple kicks result in a total accumulated momentum 
$Qs=\hat{q} \cdot L$ where L is the medium length.
 
\begin{figure}[h]
\centering
\includegraphics[width=0.43\textwidth]{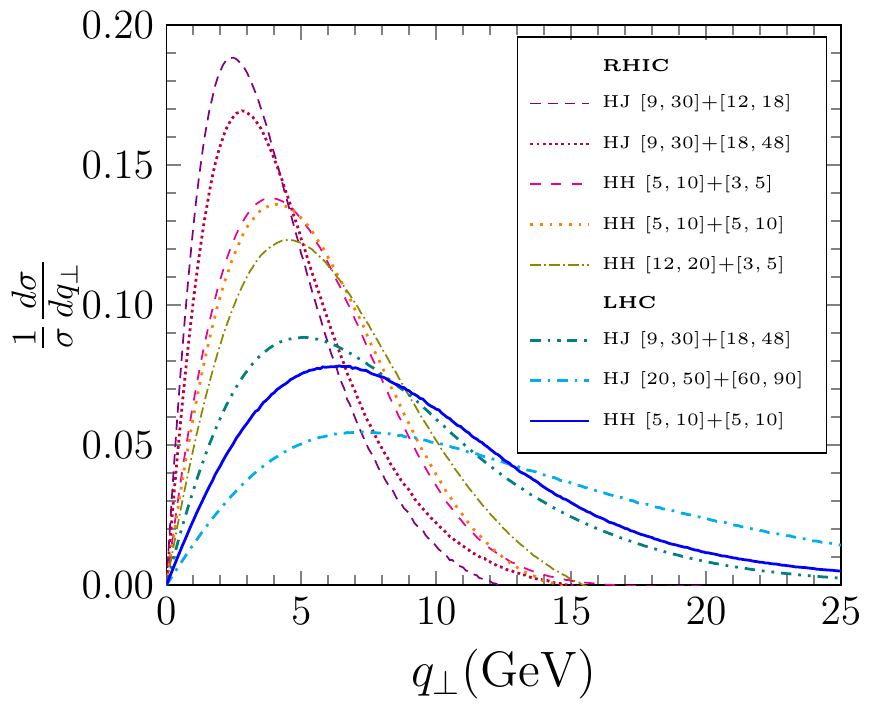}

\caption{Vacuum Sudakov contribution to momentum broadening \cite{Chen:2016vem}}
\label{fig:Sudakov}
\end{figure}

\begin{figure}[h]
\centering
\includegraphics[width=0.3\textwidth]{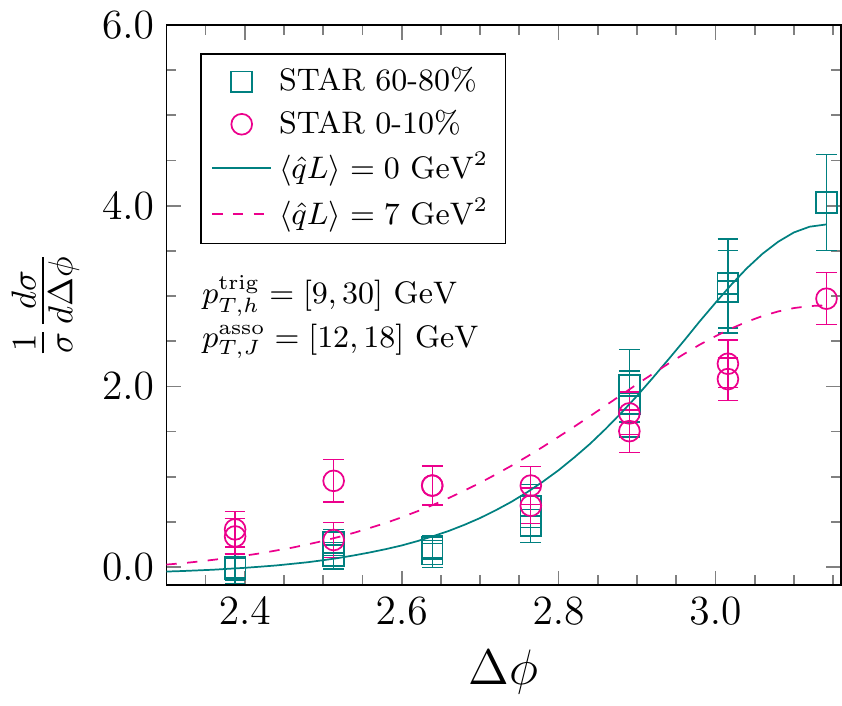}
\includegraphics[width=0.3\textwidth]{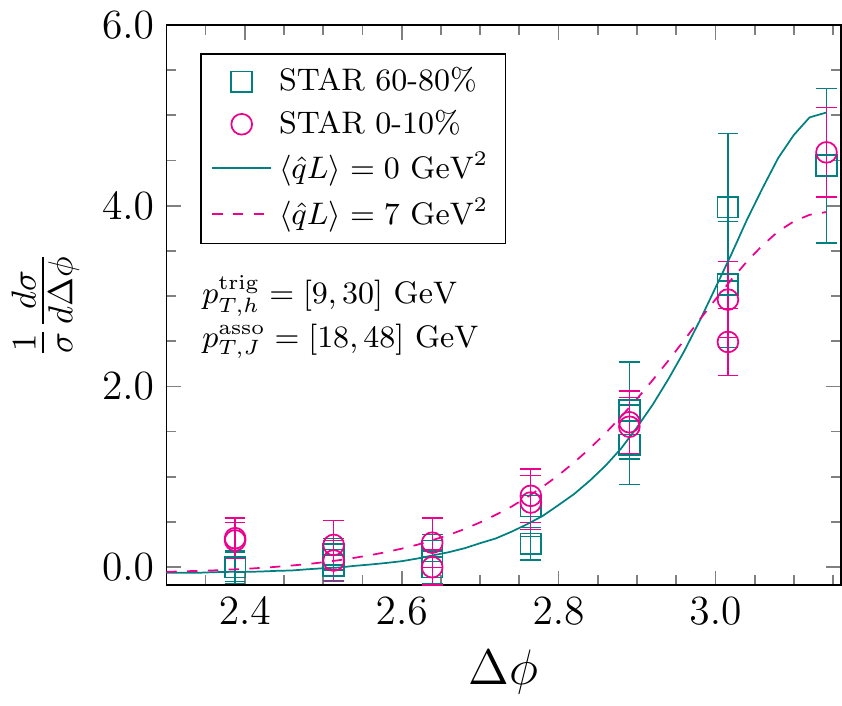}
\includegraphics[width=0.3\textwidth]{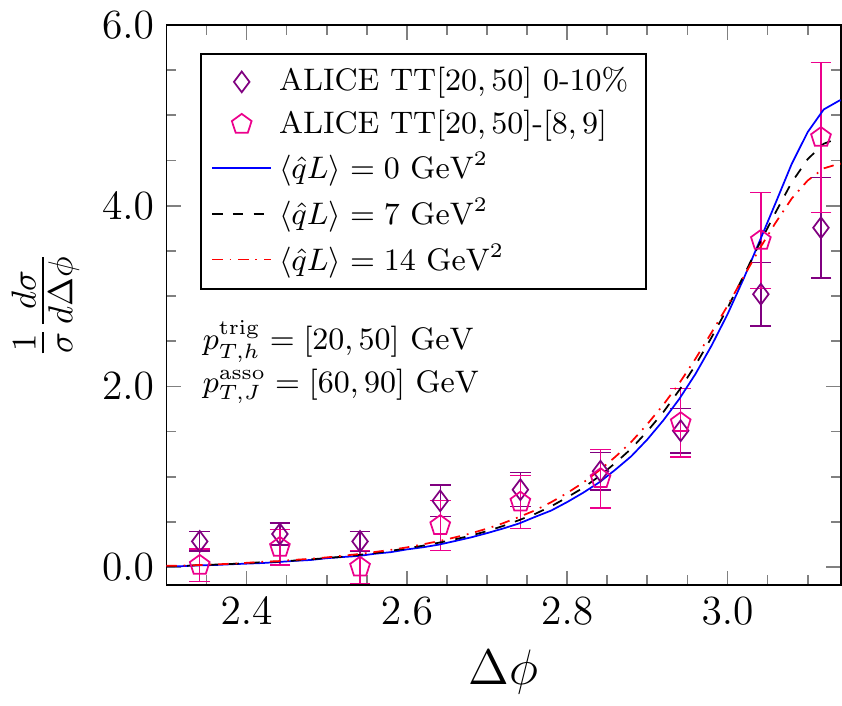}
\caption{Hadron-jet azimuthal angular correlations measured by STAR
  and ALICE compared to calculations including both vacuum and
  medium-induced azimuthal broadening \cite{Chen:2016vem} }
\label{fig:azimuthalFit}
\end{figure}

The relative contributions of vacuum and in-medium effects to the decorrelation measured at RHIC and LHC 
has being studied theoretically using resummation techniques \cite{Chen:2016vem}.
Fig. \ref{fig:Sudakov} shows the vacuum contribution to momentum
broadening for the different kinematic choices of trigger hadrons and
recoil hadrons/jets in hadron-hadron and hadron-jet correlations at
RHIC and LHC. One can see that at the LHC kinematics the vacuum distributions
are broader than at RHIC and thus medium-induced broadening, which
would be combined in quadrature with these distributions,  would be
more difficult to measure. 
Fig. \ref{fig:azimuthalFit} shows the hadron-jet angular correlations
measured at RHIC \cite{Adamczyk:2017yhe} and at LHC \cite{Adam:2015doa} for different kinematic choices for the trigger and
recoil jet momentum, compared to the calculations that also
incorporate medium-broadening, parameterized as $\langle \hat{q}L \rangle$.  One can see that as kinematic cuts select
harder scatterings, the sensitivity to different choices of broadening parameter
is decreased. 

\begin{figure}[h]
\centering
\includegraphics[width=0.4\textwidth]{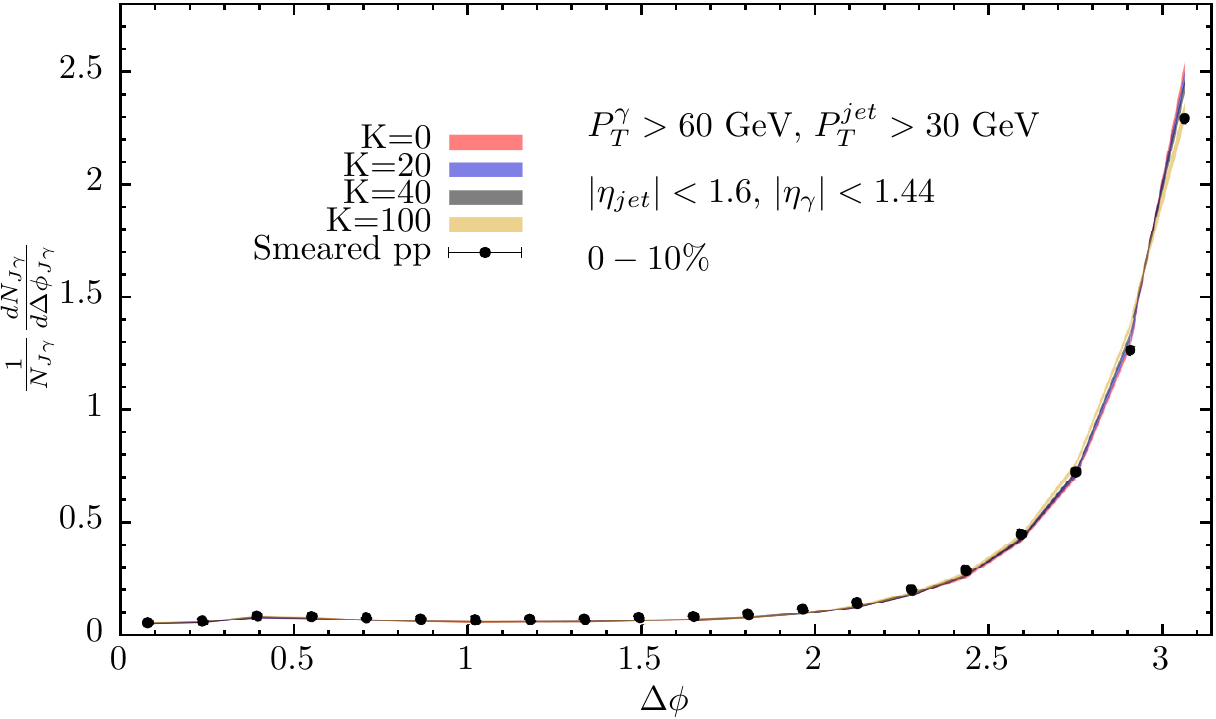}
\includegraphics[width=0.35\textwidth]{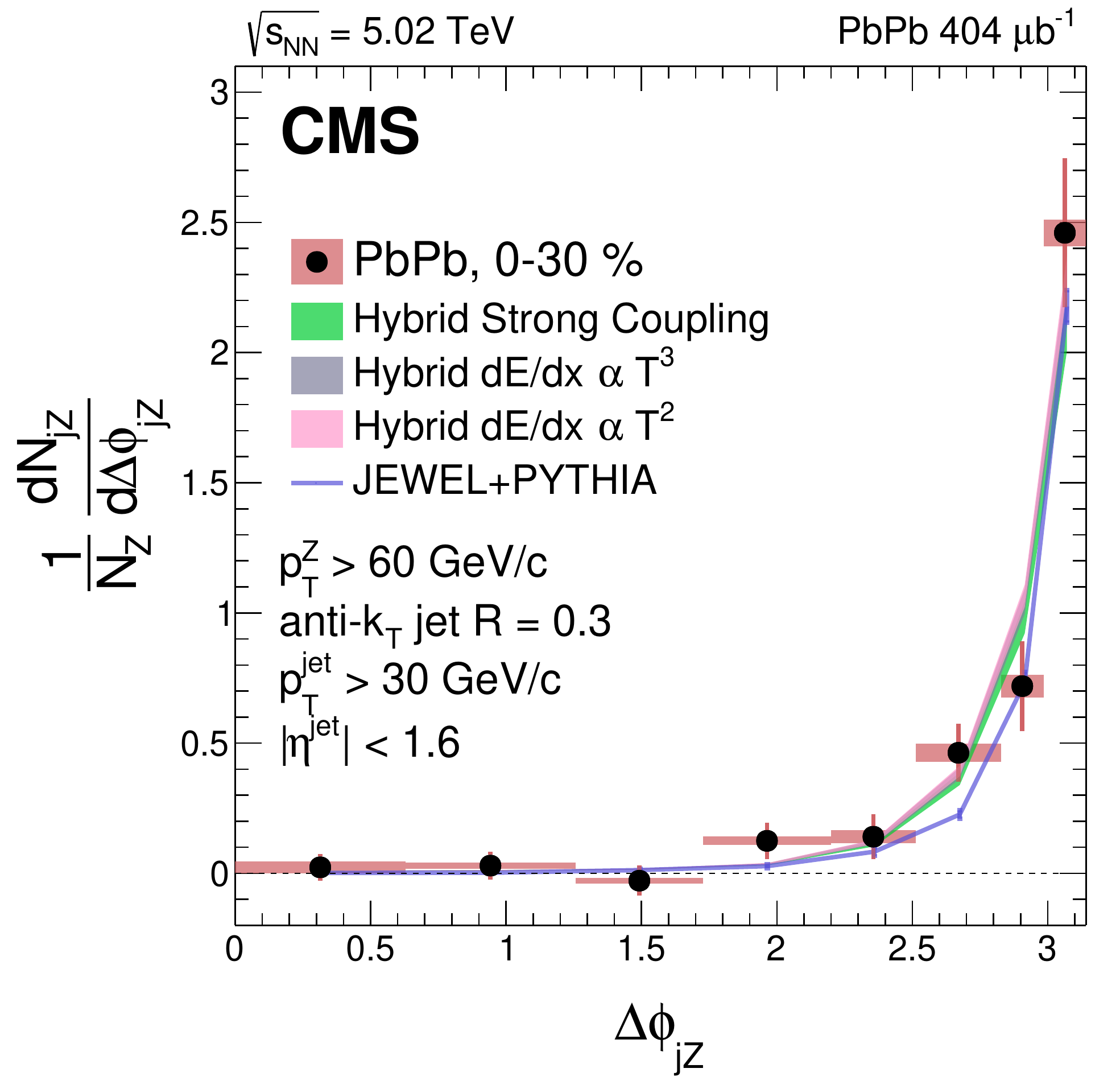}
\caption{Left: dependence of the $\gamma$-jet azimuthal correlations
  on the broadening paramter calculated with the hybrid model
  \cite{Casalderrey-Solana:2014wca} . Right: CMS $Z$-jet azimuthal
  angular correlation compared to models with different assumptions on the energy loss mechanism}
\label{fig:hybridcorrelation}
\end{figure}

 In the strongly coupled limit
\cite{Casalderrey-Solana:2014wca} there is no notion of scattering
centers or of multiple discrete scatterings. However, under
some limits, coloured excitations acquire transverse momentum
according to a gaussian with width $Q^{2}=\hat{q}L$ where $\hat{q} =
K T^{3}$ and $K$ is a free parameter of the theory. 
In Fig. \ref{fig:hybridcorrelation} (left), one can see that very different choices for the broadening parameter lead to negligible changes in the azimuthal
correlation for the kinematics selected by the CMS
cuts. Fig. \ref{fig:hybridcorrelation} (right) shows the azimuthal
angular correlation for $Z$-jet pairs
compared to purely perturbative JEWEL model and 
 to the hybrid model with different assumptions for the energy loss
 rate. The four curves can describe the data and the conclusion is that
 this observable, with the given kinematic cuts, is not very sensitive to the
 details of the microscopic dynamics of the interaction with the
 medium. 

\subsection{Tails of the azimuthal correlation}
\begin{figure}[h]
\centering
\includegraphics[width=0.45\textwidth]{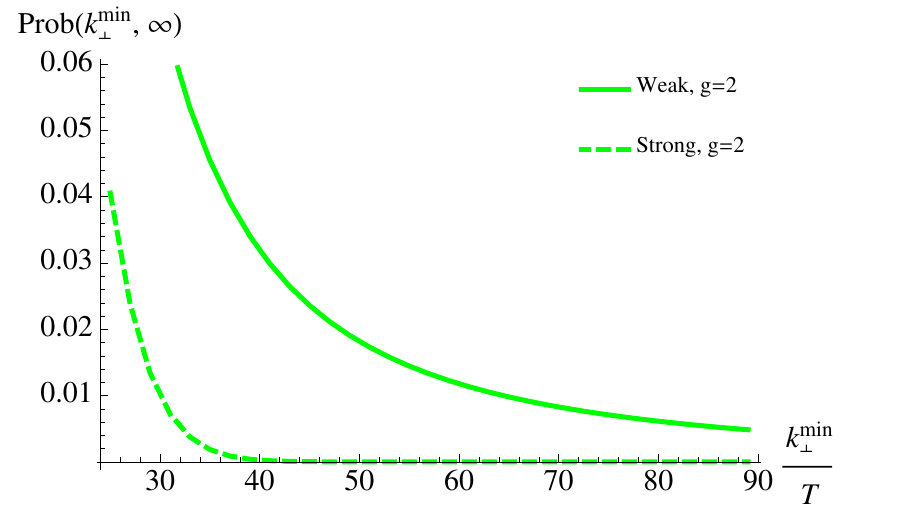}
\includegraphics[width=0.45\textwidth]{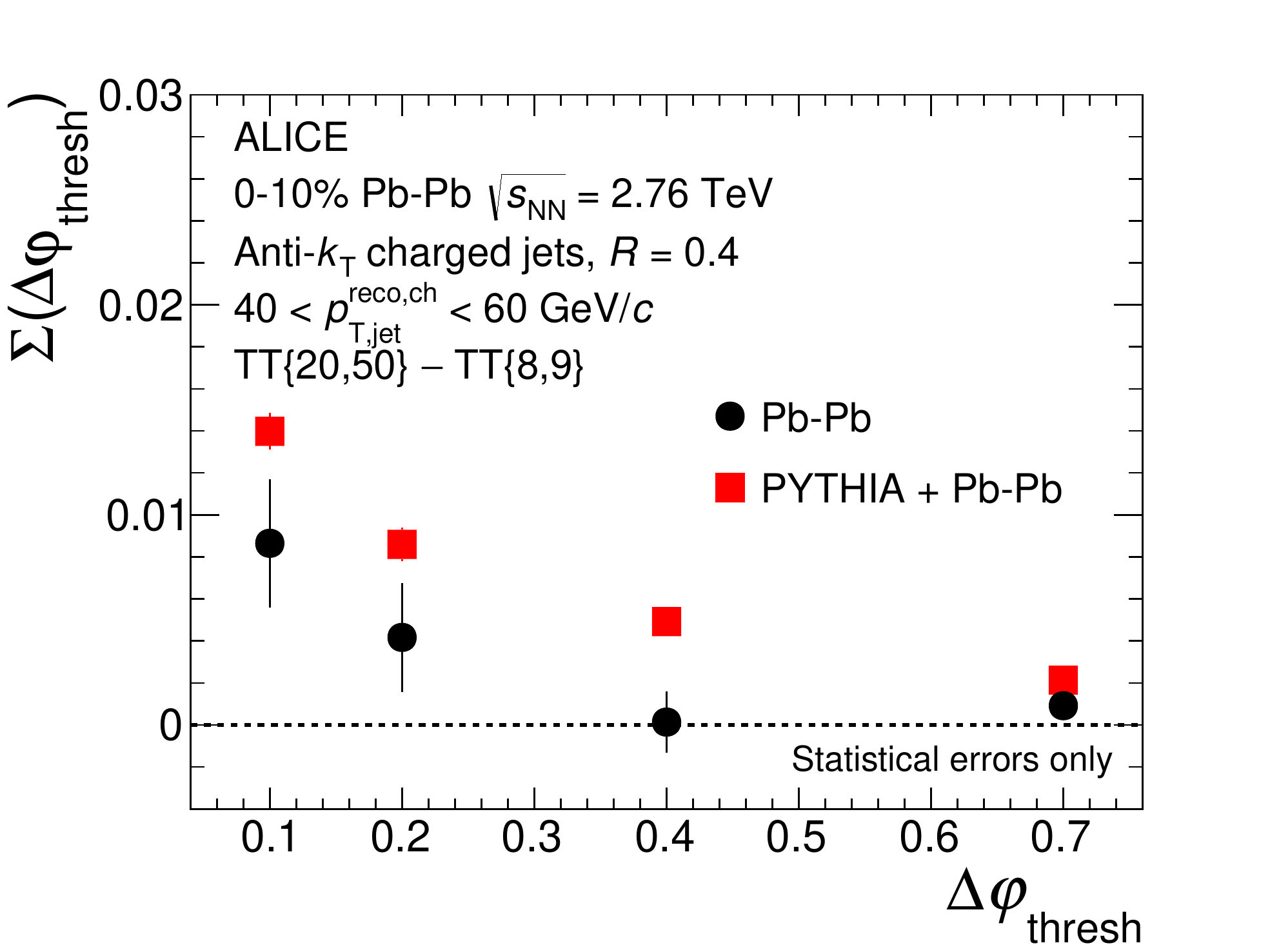}
\caption{Left: Probability to receive a transverse momentum kick from
  the medium above $k_{T}^{min}$ in the weakly and strongly coupled
  limits \cite{DEramo:2012uzl}.  Right: Integrated yield of the hadron-jet azimuthal angular
  correlation above threshold angle, by ALICE}
\label{fig:deramo}
\end{figure}

The main peak of the azimuthal angular correlation helps to constrain the
average momentum broadening while the tails of such distribution might encode
fundamental information on the dynamics of the degrees of freedom of
the medium.  
Recent calculations \cite{DEramo:2012uzl} show that the probability of large angle and
semi-hard parton deflections in medium, the so-called Moliere regime,
is parametrically larger in the case of quasi-particle than in the
case of strongly-coupled degrees of freedom. This is shown in
Fig. \ref{fig:deramo} (left).  The calculations are done in
terms of transverse momentum $k_{T}$ and in the limit of
infinite parton energy. More realistic calculations will open a very
interesting possibility of measuring an excess of large angle
deflections in heavy ion collisions relative to the vacuum reference
as a signature of quasi-particle degrees of freedom. The inspection of
the tails of the azimuthal correlation requires large statistics and
good control of the background and of the higher order contributions
needed to describe that range already in vacuum. ALICE \cite{Adam:2015doa} has
measured the integrated yield above azimuthal angle threshold in
hadron-jet correlations in Pb-Pb collisions. Fig. \ref{fig:deramo}
(right) shows the integrated yield above azimuthal angle threshold
$\Delta \varphi_{thresh}$ in the hadron-jet correlation both for 
data and the smeared reference for pp collisions.  
No significant change of trend of data relative to the smeared 
reference with angular threshold is observed. However, note that the recoil jet
$p_{T}$ has momentum above 40 GeV and that low $p_{T}$ is preferable,
since the deflection angle resulting from the given momentum from the medium decreases with jet energy.

Unfolding in 2D  will help to
correct simultaneously the $p_{T}$ of the jet and the azimuthal angle
between the jet and the trigger hadron or boson, allowing for fully
corrected measurements of $\Delta_{\phi}$ at significantly smaller jet
$p_{T}$.  

\section{Conclusions}
A large sample of differential jet measurements is currently
available for systematic comparisons to the theory models. 
Most of the jet results from LHC Run 2 can be described simultaneusly by
weakly coupled and strongly coupled theory models, implying lack of sensitivity to discriminate among
different microscopic pictures for jet energy loss. However, several of
the experimental measurements have strong cuts on jet $p_{T}$ that
might be biased to a kinematic regime dominated by vacuum effects, whereby
medium effects become a small perturbation that is difficult to
disentangle. 
The access to low jet $p_{T}$  and to more differential intrajet
measurements will open new possibilities in the data-theory
comparison.   
\section{Acknowledgement}
 I thank the organisers for the interesting conference and the opportunity to give this talk, and Peter Jacobs, Christian Klein-Boesing and Marco van
Leeuwen for discussions and critical reading of this manuscript.


\bibliographystyle{elsarticle-num}
\bibliography{<your-bib-database>}









\end{document}